\documentclass[aps,prb,showpacs,twocolumn,superscriptaddress]{revtex4-1}

\usepackage{graphicx}
\usepackage{amsmath}
\usepackage{amssymb,gensymb}
\usepackage{wasysym}
\usepackage{textcomp}
\usepackage{color}

\begin{document}

\title{Stability of edge magnetism against disorder in zigzag MoS$_2$ nanoribbons}
\author{P\'eter Vancs\'o }
\affiliation{2D Nanoelectronics "Lend\"ulet" Research Group, Institute  of Technical Physics and 
Materials Science, HAS Centre for Energy Research, 1121 Budapest, Hungary}
\affiliation{Department of Physics, 
University of Namur, 61 rue de Bruxelles, 5000 Namur, Belgium}
\author{Imre Hagym\'asi}
\affiliation{Department of Physics and Arnold Sommerfeld Center for Theoretical Physics, 
Ludwig-Maximilians-Universit\"at M\"unchen, D-80333 M\"unchen, Germany}
\affiliation{Strongly Correlated Systems "Lend\"ulet" Research Group, Institute for Solid State Physics and Optics, HAS Wigner Research Centre
for Physics, 1121 Budapest, Hungary}
\author{Pauline Castenetto}
\affiliation{Department of Physics, 
University of Namur, 61 rue de Bruxelles, 5000 Namur, Belgium}
\author{Philippe Lambin}
\affiliation{Department of Physics, 
University of Namur, 61 rue de Bruxelles, 5000 Namur, Belgium}

\begin{abstract}
Molybdenum disulfide nanoribbons with zigzag edges show ferromagnetic and metallic properties based on previous \emph{ab-initio} calculations. The investigation of the role of disorder on the magnetic properties is, however, still lacking due to the computational costs of these methods. In this work we fill this gap by studying the magnetic and electronic properties of several nanometer long MoS$_2$ zigzag nanoribbons using tight-binding and Hubbard Hamiltonians. Our results reveal that proper tight-binding parameters for the edge atoms are crucial to obtain quantitatively the metallic states and the magnetic properties of MoS$_2$ nanoribbons. With the help of the fine-tuned parameters, we perform large-scale calculations and predict the spin domain-wall energy along the edges, which is found to be significantly lower compared to that of the zigzag graphene nanoribbons. The tight-binding approach allows us to address the effect of edge disorder on the magnetic properties. Our results open the way for investigating electron-electron effects in realistic-size nanoribbon devices in MoS$_2$ and also provide valuable information for spintronic applications.
\end{abstract}
\maketitle
\section{Introduction} 
 Electronic and spintronic applications of two dimensional (2D) materials are in the focus of scientific attention. \cite{c_1, c_2,c_3,c_4,c_5,c_6} Transition metal dichalcogenides (TMDs), and particularly molybdenum disulfide (MoS$_2$), are one of the most intensively studied materials due their direct band gaps,\cite{c_4, c_7} which make them good candidates for optical\cite{c_8} and electronic applications such as transistors\cite{c_9} or even microprocessors.\cite{c_10} MoS$_2$ is a nonmagnetic semiconductor, however, several theoretical studies\cite{c_11, c_12,c_13} reported magnetic moments on the edges of zigzag MoS$_2$ nanoribbons similar to the case of zigzag graphene nanoribbons.\cite{c_14, c_15} These density functional theory (DFT) calculations revealed ferromagnetic and metallic behavior of the edges, furthermore the emerged magnetism was still preserved in several edge reconstructed and passivated systems independent from the nanoribbon’s width.\cite{c_11, c_16,c_17,c_18} In the last years, impressive advances on the sample preparation were reported. MoS$_2$ nanoribbons with nanometer width and well-defined edges have been synthetized by using bottom-up\cite{c_19, c_20} and top-down\cite{c_21, c_22} techniques. In addition, magnetic measurements on large scale epitaxial growth of zigzag MoS$_2$ nanoribbons show prominent ferromagnetic behavior.\cite{c_23} Due to these recent experimental results theoretical understanding of the edge magnetism in MoS$_2$ nanoribbons is quite important, especially in larger, realistic systems including disorder.
\par In the case of zigzag graphene nanoribbons different spintronic applications were proposed based on the appeared magnetic moments along the edges.\cite{c_24, c_25,c_26,c_27} However, it has turned out that the computational cost of \emph{ab-initio} calculations does not allow the investigation of graphene nanoribbons in realistic size with disorder. In order to study systems involving a large number of atoms, tight-binding (TB) approach is a more suitable alternative, which can also provide a simple starting point for the further inclusion of many-body electron-electron effects. By using TB parameters and local Coulomb interaction (the so-called Hubbard-$U$) the magnetic properties of the graphene nanoribbons were studied in large scale systems.\cite{c_28, c_29,c_30,c_31} It was found that the magnetism of the edge states is robust against disorder and potential fluctuations. However, as far as we know, similar investigation of the magnetic properties of MoS$_2$ nanoribbons has not been performed yet.
\par Nowadays, a wide range of TB parameters is available for TMDs including MoS$_2$ monolayers.\cite{c_32, c_33,c_34,c_35} These models accurately reproduce the DFT band structure calculations near the conduction and valence bands, providing a key tool for further studies of electronic and transport properties in larger systems. In the case of MoS$_2$ nanoribbons several papers examined the electronic and transport properties based on the monolayer MoS$_2$ TB parameters.\cite{c_36, c_37,c_38,c_39,c_40,c_41} Besides the important observations of these works, none of them takes into account the different environment of the edge atoms compared to the inner atoms of the nanoribbon during the TB parametrization. Therefore, the obtained band structures of the nanoribbons, where all of the Mo and S atoms are handled equally, show only qualitative agreement with the DFT band structure calculations. Namely, the TB calculations display metallic properties of the zigzag nanoribbons, but even the number of the metallic bands is different compared to the DFT results. However, in order to describe the proper magnetic properties of the nanoribbons, the accurate treatment of the edge states is crucial, since the magnetism originates from the splitting of the metallic edge states.
\par In this paper we demonstrate that the metallic states of zigzag MoS$_2$ nanoribbons can be reproduced with their proper orbital characters by adapting TB parameters for the edge atoms. These results provide us a starting point for further inclusion of the electron-electron interactions, where the Coulomb repulsion is taken into account by using local Hubbard interaction terms in the fine-tuned TB Hamiltonian. This Hubbard model within the mean-field approximation, applied for the first time to MoS$_2$ nanoribbons, circumvents the computational bottleneck of \emph{ab-initio} calculations. Our results show that this simple model is not only capable of describing the magnetism in MoS$_2$ nanoribbons with zigzag edges, but also gives quantitatively accurate results for the magnetization values compared to DFT calculations. As a next step, we extend our calculations to several nanometer long ribbon containing 800 atoms. We have found that the domain-wall energies are much lower compared to those of the graphene nanoribbons,\cite{c_42} which predicts fluctuations of spins along the ribbon edge. We also investigated short- and long-range disorder originating from inhomogeneous charge distribution of the substrate or other structural imperfections.
\par The paper is organized as follows. In Sec.~II.~we present the applied theoretical models for MoS$_2$ nanoribbons with zigzag edges. In Sec.~III.~A we compare the TB and DFT calculations of the electronic and magnetic properties of the nanoribbons. In Sec.~III.~B we applied our method for several nanometer long nanoribbons and analyze the spin domain wall. Sec.~III.~C presents the effects of the short and long-range disorder on the magnetic properties. Finally, we show our conclusions in Sec.~IV.
\section{Methods} 
\begin{figure}[!ht]
\includegraphics[width=\columnwidth]{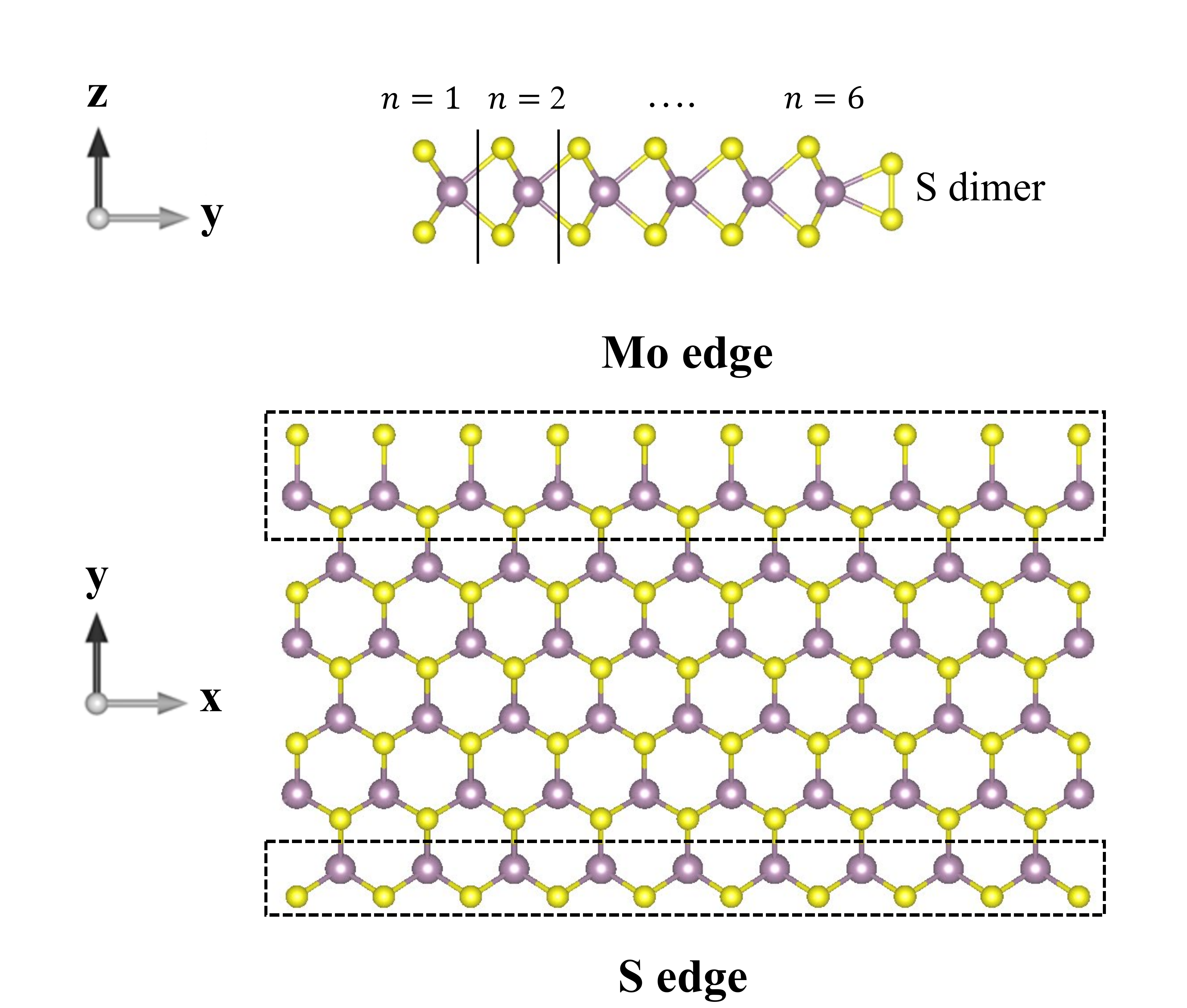}
\caption{Side and top view of the relaxed geometry of zigzag MoS$_2$ nanoribbon with sulfur dimer passivation along the Mo edge. Purple and yellow spheres represent molybdenum and sulfur atoms, respectively. The edge regions, where TB parameters of the atoms are modified marked by the dotted black lines.}
\label{fig:ribbon_geometry}
\end{figure}

Band structure calculations of the zigzag MoS$_{2}$  nanoribbon (Fig.~\ref{fig:ribbon_geometry}) are performed by using DFT and TB calculations. As Capuletti et al.\cite{c_33} pointed out an eleven-orbital TB model within Slater-Koster scheme \cite{c_43} is able to reproduce the band structure of the single layer MoS$_{2}$. This model considers an orthogonal basis composed of five orbitals (4$d_{xy}$, 4$d_{yz}$, 4$d_{xz}$, 4$d_{x^2-y^2}$, 4$d_{3z^2-r^2}$) for each molybdenum (Mo) atom and three orbitals (3$p_x$, 3$p_y$, 3$p_z$) for each sulfur (S) atom resulting in $z$-symmetric and $z$-antisymmetric states. 
In the case of our zigzag nanoribbon calculations, we follow the method and use the TB parameters (hopping terms and on-site energies) described in Ref.~[\onlinecite{c_33}]  as a starting point. In the next step, we modify the on-site energy parameters of the edge atoms in the ribbon (Table~\ref{table:parameters}) in order to quantitatively reproduce the DFT band structure results.
\par DFT calculations are carried out using the projector augmented wave (PAW) method\cite{c_44} as implemented in the Vienna ab initio simulation package (VASP).\cite{c_45} The generalized gradient approximation of Perdew--Burke--Ernzerhof (GGA-PBE) is adopted for the exchange-correlation (XC) functional.\cite{c_46} The band structure calculations are performed with plane wave cutoff of 500 eV and the Brillouin zone is sampled with ($12\times12\times1$) Monkhorst-Pack mesh of $k$-points.\cite{c_47} During geometry optimization, the convergence criterion for forces is set to 0.01 eV/\AA.
\begin{figure*}[!ht]
\includegraphics[scale=0.185]{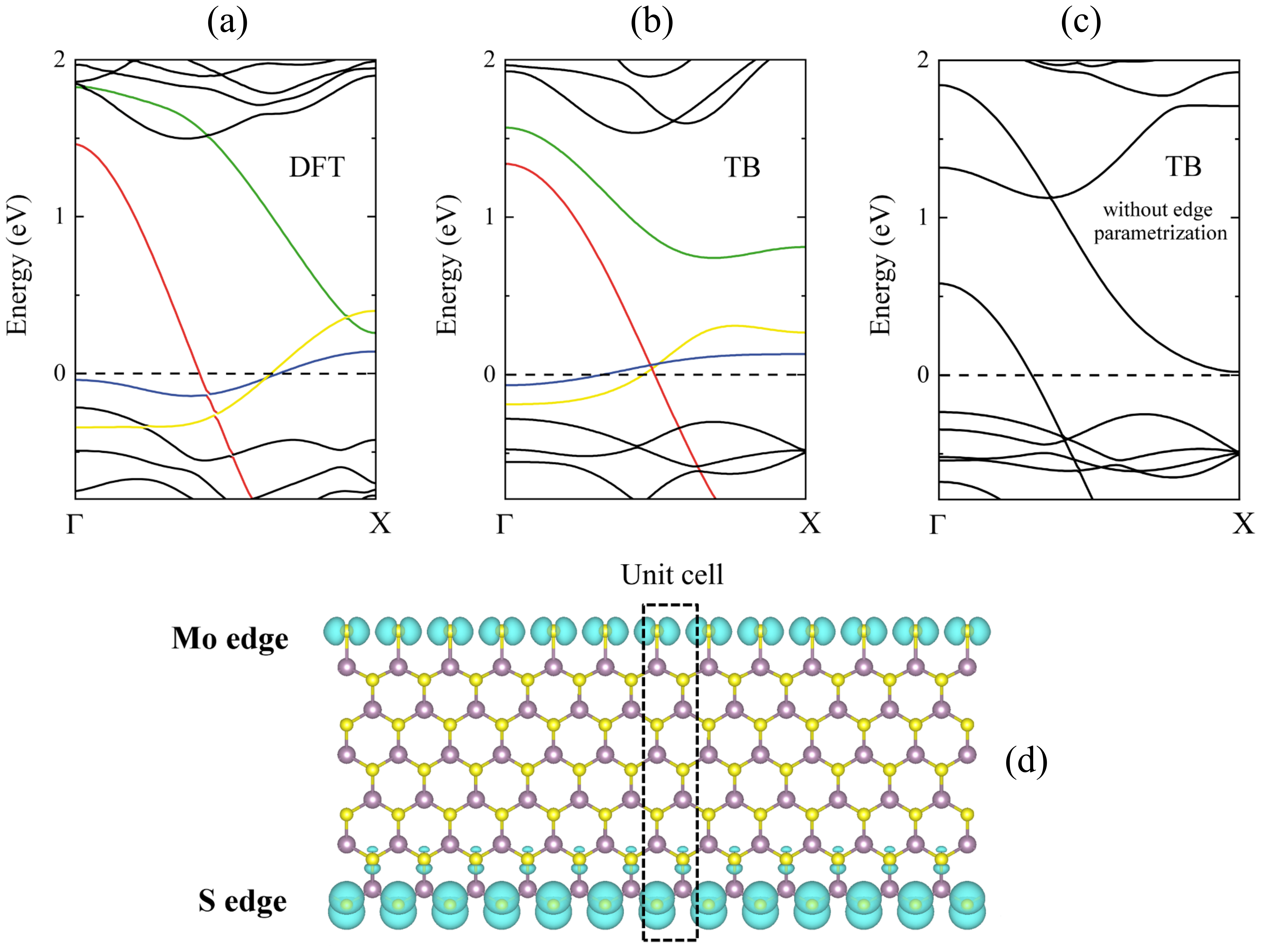}
\caption{Non spin-polarized band structure calculations of the zigzag MoS$_2$ nanoribbon ($n=6$). (a) DFT calculation, (b) TB calculation using the modified TB parameters for the edge atoms. Mid-gap states originated from the edge atoms are marked by different colors. States with blue and green colors correspond to the S edge, while red and yellow colors correspond to the Mo edge. (c) TB calculations without edge parametrization. (d) Charge density plots around the Fermi-level calculated by DFT. The isovalue is set to be $2\times10^{-3}$ e/\AA$^3$. }
\label{fig:ribbon_bands}
\end{figure*}
\par To describe the magnetic properties of the zigzag nanoribbons with the help of the modified TB parameters we use a grand-canonical ensemble and switch on Hubbard interaction terms with different amplitudes, $U_{\rm Mo}$, $U_{\rm S}$ corresponding to the five Mo and three S orbitals, respectively:
\begin{equation}
\begin{split}
\label{eq:Hubbard}
 \mathcal{H}=\sum_{\langle ij
\rangle\sigma}t_{ij}\hat{c}^{\dagger}_{i\sigma}\hat{c}^{\phantom\dagger}_{j\sigma}+U_{\rm Mo}\sum_{i\in{\rm Mo}}\hat {
n } _ {i\uparrow}
 \hat{n}_{i\downarrow} \\
+U_{\rm S}\sum_{i\in{\rm S}}\hat {n } _ {i\uparrow}
 \hat{n}_{i\downarrow} 
-\sum_{i\sigma}(\varepsilon_i-\mu)\hat {n }_{i\sigma}.
\end{split}
\end{equation}
Here $t_{ij}$ encodes the hopping TB parameters, $c_{i\sigma}$ annihilates a fermion at site $i$ with spin $\sigma$, $\varepsilon_i$ are the fine-tuned on-site energy parameters, $n_{i\sigma}$ is the particle-number operator, and $\mu$ is the chemical potential. The summation in the two Hubbard terms extends over the Mo or S sites only. In the case of the long nanoribbons our system consists of 800 atoms and nearly 3000 fermionic sites, at a rate of one site per atomic orbital. Such a large system can be solved only using some kind of approximation. We apply the standard mean-field decoupling of the Hubbard terms:
${n}_{i\uparrow}{n}_{i\downarrow}\approx \langle {n}_{i\uparrow}\rangle{n}_{i\downarrow}+{n}_{i\uparrow}\langle{n}_{i\downarrow}\rangle-
\langle{n}_{i\uparrow}\rangle\langle{n}_{i\downarrow}\rangle$ which gives us an effective single-particle Hamiltonian that can be diagonalized either in $k$- or real space self-consistently. The chemical potential is also determined in each iteration step by requiring that the electron number per unit cell should give the same number as in the nonmagnetic case. The iteration is stopped if the difference between the electron densities decreases below $10^{-6}$.
\section{Numerical results and discussion}
\subsection{Band structure calculations of the nanoribbon}
In Fig.~\ref{fig:ribbon_geometry} we can see the top and side view of the geometry of MoS$_2$ nanoribbon ($n=6$) with zigzag edges. 
The width parameter $n$ is defined as the number of 
zigzag lines across the nanoribbon as defined analogously for the case of graphene nanoribbons. Without edge passivation, zigzag nanoribbons have two types of edges, one is S-terminated, while the opposite one is Mo-terminated. However, both theoretical predictions and experimental observations found the pure Mo edge energetically unfavoured compared to edge passivated geometries.\cite{c_48,c_49} In order to model realistic nanoribbon geometries in our calculations, we use the experimentally observed sulfur dimers passivation at the Mo edge. After the relaxation of the geometry, the S-S bond length of the dimer is found to be 1.99 \AA, significantly differ from the in-plane (3.18 \AA) and out-of-plane (3.13 \AA) S-S bond lengths.
\par
\begin{figure}[!ht]
\includegraphics[width=\columnwidth]{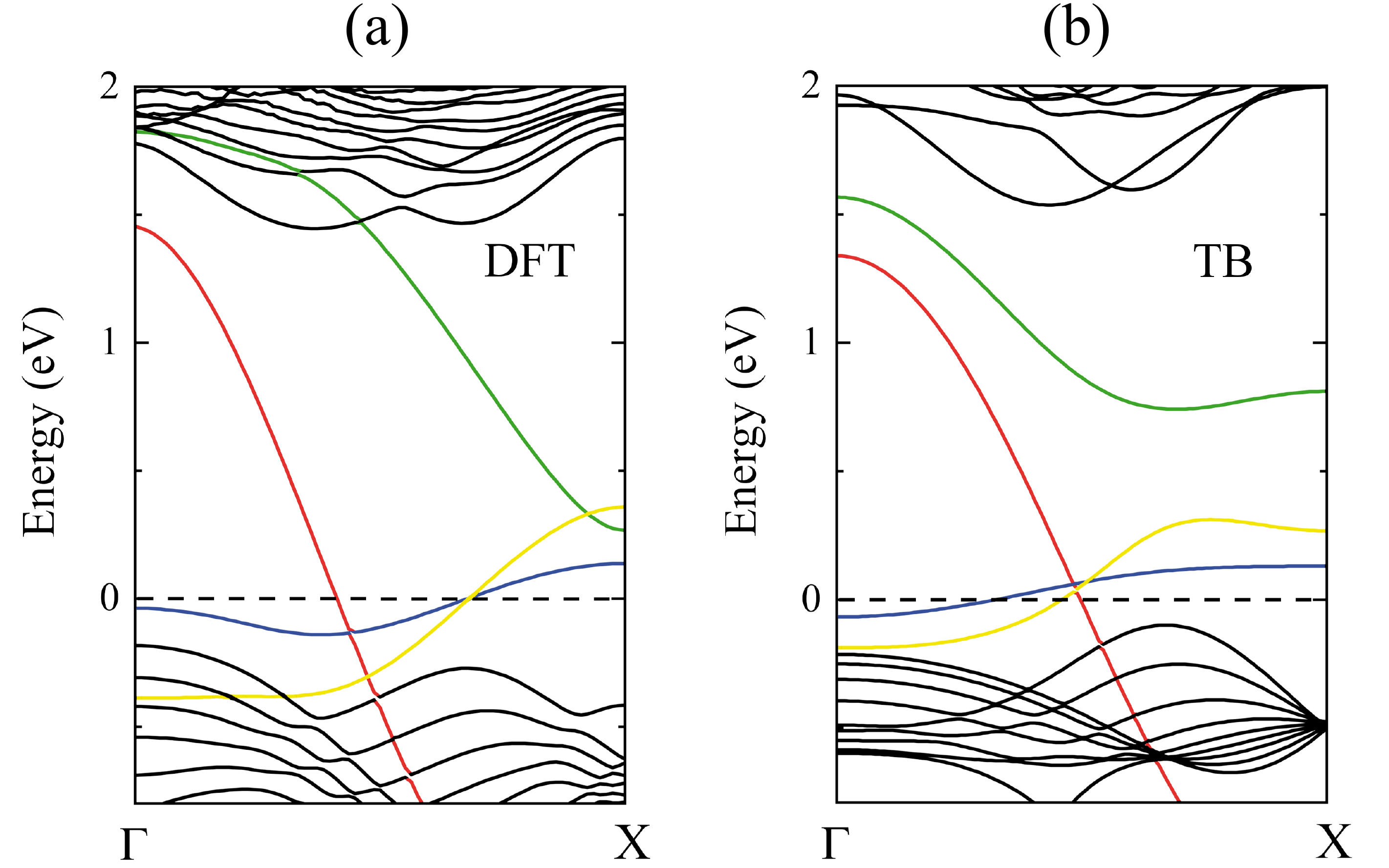}
\caption{Non spin-polarized band structure calculations of the zigzag MoS$_2$ nanoribbon ($n=12$). (a) DFT calculation (b) TB calculation using the modified TB parameters for the edge atoms. Mid-gap states (marked by different colors) are the same compared to narrower ($n=6$) nanoribbon implying negligible electronic interaction between the Mo and S edges.}
\label{fig:ribbon_bands_12}
\end{figure}

 The unit cell band structure calculations of the nanoribbon ($n=6$) without spin polarization are shown in Fig.~\ref{fig:ribbon_bands}. From the DFT calculations (Fig.~\ref{fig:ribbon_bands}(a)) four mid-gap states can be seen that are highlighted by different colors. Three of them cross the Fermi level, implying the existence of metallic states in agreement with previous results.\cite{c_48,c_50} These metallic states are almost completely localized on the S and Mo edge of the nanoribbon (Fig.~\ref{fig:ribbon_bands}(d)). More precisely, the Kohn-Sham wave functions reveal that the states marked by blue and green colors correspond to the localized states of the S and Mo atoms at the S edge, while the states with red and yellow colors originate from S dimers and Mo atoms at the Mo edge. From the charge density plot around the Fermi-level (Fig.~\ref{fig:ribbon_bands}(d)) it is also visible that in the Mo edge side the S dimers have $p_x$ orbital character in contrast to the S edge side, where the $p_z$ orbitals of the S atoms dominate, forming one-dimensional metallic states along the edges.
\begingroup
\squeezetable
\begin{table}[!h]
\centering 
\begin{tabular}{c |c| c| c}
\hline\hline 
\textbf{-} & \textbf{Mo atom} & \textbf{S atom} & \textbf{S dimer} \\ [0.5ex] 
\hline 
\textbf{Mo edge} & -2.03, 1.42, 1.42,  & 0.28,-8.28, -12.24 & -0.55, -5.28,-8.24 \\ 
 & -4.03, 0.51 & & \\\hline
\textbf{S edge} & -2.03, 4.30, -0.80, & -1.90, 0.18, -6.50 & - \\ [1ex] 
 & -12.03, -2.60 & & \\
\hline\hline
\end{tabular}
\caption{Modified tight-binding on-site energy parameters for the edge atoms (marked by the dotted black lines in Fig.~\ref{fig:ribbon_geometry}) in units of eV. Values corresponds to the five orbitals ($d_{xy}$, $d_{yz}$, $d_{xz}$, $d_{x^2-y^2}$, $d_{3z^2-r^2}$) for Mo atoms, and three orbitals ($p_x$, $p_y$, $p_z$) for S atoms, respectively. On-site energies and the hopping terms for the inner atoms of the nanoribbons are given in Ref.~[\onlinecite{c_33}].}
\label{table:parameters}
\end{table}
\endgroup
By exploiting our modified TB parameters (Table~\ref{table:parameters}) in the edge regions (Fig.~\ref{fig:ribbon_geometry} dotted area), we are able to reproduce the shape and the number of the metallic states with their proper orbital character (Fig.~\ref{fig:ribbon_bands}(b)). In contrast, the band structure results without our fine-tuning for the edge atoms (Fig.~\ref{fig:ribbon_bands}(c)) show significantly different edge states compared to the DFT results highlighting the importance of the proper treatment of the edges within the TB formalism. We have also examined the electronic interaction between the two edges of the nanoribbon. We perform the same calculations on a double size ($n=12$) nanoribbon (Fig.~\ref{fig:ribbon_bands_12}), where the mid-gap states show exactly the same behavior. Our results have verified that the edges states can be treated independently even in the case of the narrower ($n=6$) nanoribbon. In other words, the electronic states at the S edge and Mo edge do not interact with each other for nanoribbons having $n\geq6$, which implies that our TB parametrization of the edges is able to describe wider ribbons electronic properties as well.
\par
\begin{figure}[!ht]
\includegraphics[width=\columnwidth]{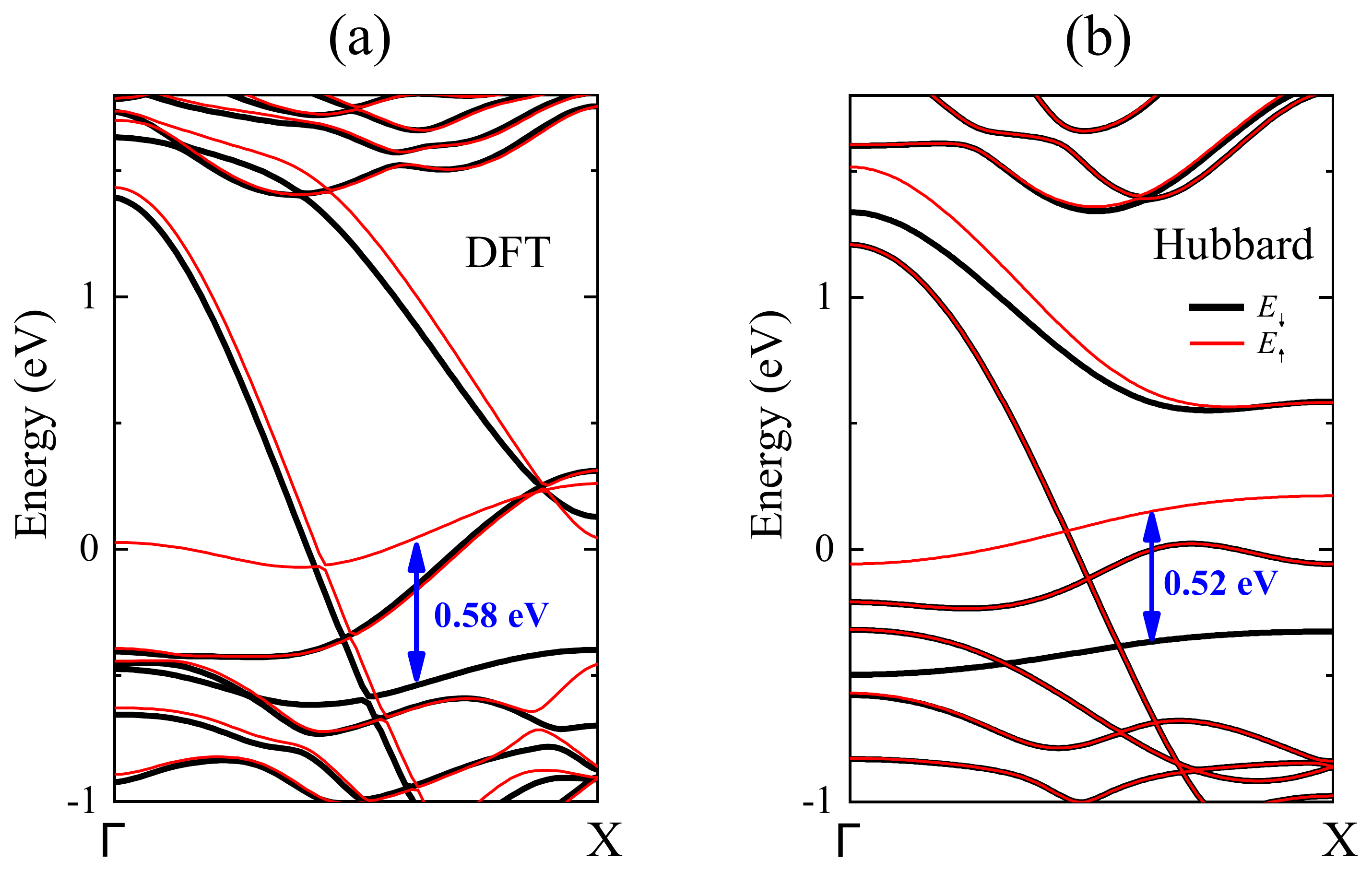}
\caption{Spin-polarized band structure calculations of the zigzag MoS$_2$ nanoribbon ($n=6$) by using (a) DFT and (b) Hubbard-model with $U_{\rm S}=1.7$ eV and $U_{\rm Mo}=0.6$ eV parameters. Red and black curves correspond to the up and down spins, respectively. The Fermi level is at zero energy. Magnetization appears from the splitting of the band of the S atom resulting localized magnetic moments along the S edge.}
\label{fig:ribbon_spin_bands}
\end{figure}

 As a next step, we apply the Hubbard-model, Eq.~(\ref{eq:Hubbard}), by using our modified TB parameters to obtain the magnetic properties of the nanoribbons. The Hubbard interaction terms for the S and Mo atoms are defined by comparing the results with spin polarized DFT calculations. 
Figure~\ref{fig:ribbon_spin_bands} presents the results of the two calculations, where $U_{\rm S}=1.7$ eV and $U_{\rm Mo}=0.6$ eV values are applied in the Hubbard calculations. 
According to both methods localized magnetic moments appear on the S edge, which result from the spin splitting of the flat band of the edge S atoms (marked by blue color in Fig.~\ref{fig:ribbon_bands} (a)-(b)). In more detailed, the higher value of density of states and the finite Coulomb repulsion leading to Stoner instability and splits the partially occupied S atom band into a totally filled spin-down ($\downarrow$) and an almost empty spin-up ($\uparrow$) band. By using the appropriate $U$ values, the large 0.52 eV splitting of the band and the local magnetic moments $M= 1/2({n}_{\uparrow}-{n}_{\downarrow}) g_e\mu_B\approx 0.35$ $\mu_B$ on the S atoms at the S edge (both in the upper and bottom layer) are predicted from the Hubbard calculations in excellent agreement with the DFT results ($M= 0.32$ $\mu_B$). 
We note that the metallic state from the S atom, which plays the major role in the magnetization, is completely missing in the non-parametrized TB results (Fig.~\ref{fig:ribbon_bands}(c)). From our spin-polarized calculations only the S edge exhibits magnetic properties, in contrast to the previous DFT calculation with unpassivated Mo edge.\cite{c_11}  The vanishing magnetic values at the Mo edge in our nanoribbon geometry are due to the S dimers passivation. Therefore, in the following sections we focus on the magnetic properties of the S edge.
\subsection{Spin domain wall}
\begin{figure*}[!ht]
\includegraphics[scale=0.333]{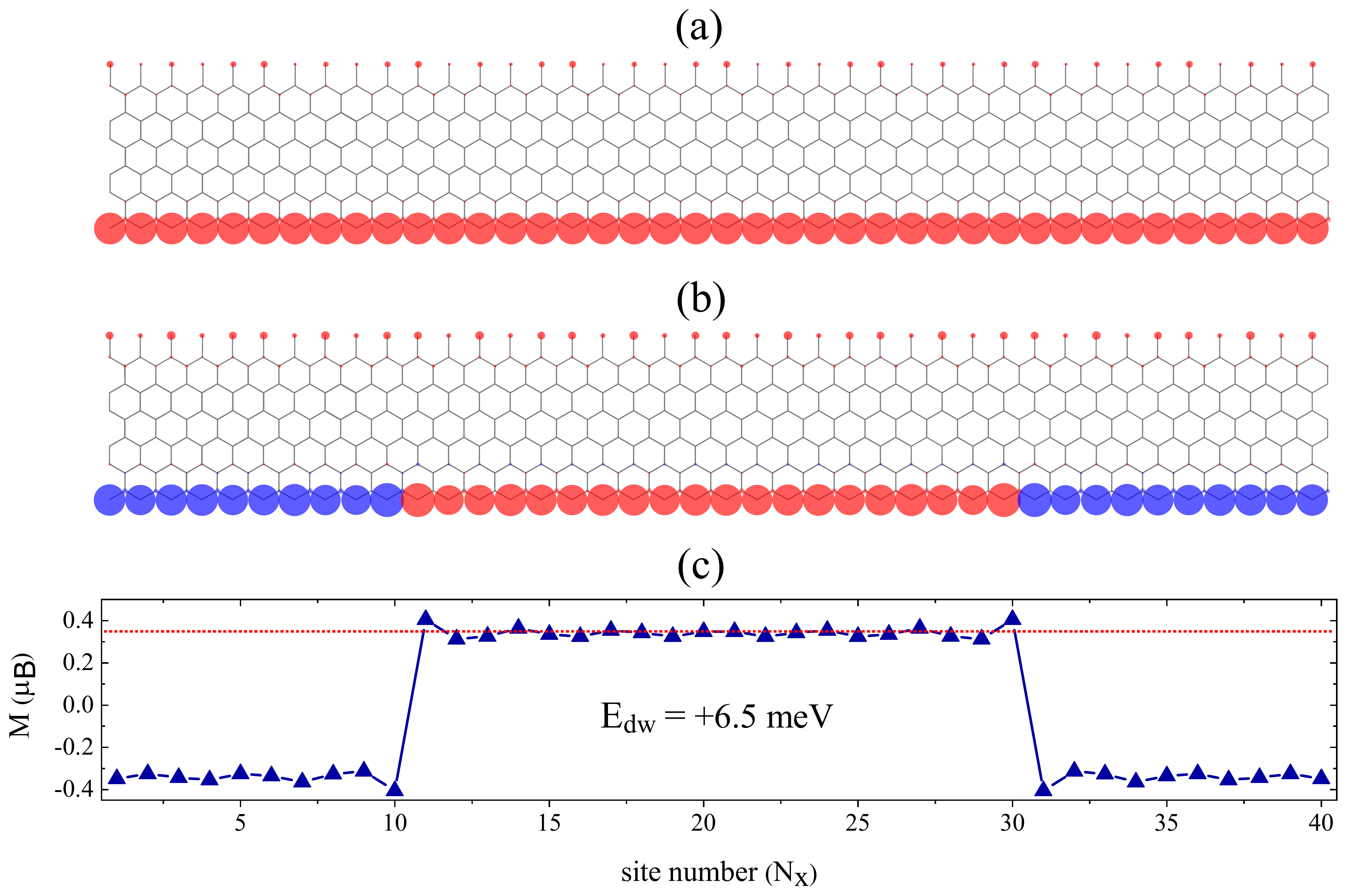}
\caption{Spin density plot of the zigzag MoS$_2$ nanoribbon. (a) Top view of the ferromagnetic ground state and (b) the collinear domain wall excitation at the S edge. Blue and red circles correspond to spin-up and spin-down electrons, respectively. (c) Magnetic moments at the edge on the S atoms in the presence of the domain wall. For comparison red dotted line shows the ground state magnetic values.}
\label{fig:ribbon_domain_wall}
\end{figure*}
\par As we demonstrated in the previous section, with the help of proper TB parametrization at the edges both the electronic and magnetic properties of a zigzag nanoribbon can be obtained. Using the results of the unit cell calculations, we extend the system size and investigate the magnetic properties of a 40 unit cell long ($L_x=12.8$ nm) nanoribbon within the framework of the Hubbard model. 
First we study collinear domain walls at the S edge by rotating the half of the spins in the supercell geometry. Fig.~\ref{fig:ribbon_domain_wall} illustrates the distribution of the spin density for the ferromagnetic ground state (Fig.~\ref{fig:ribbon_domain_wall}(a)) and the excited state including collinear domain walls (Fig.~\ref{fig:ribbon_domain_wall}(b)). The spin densities of the S atoms at the S edge (both in the upper and bottom layer) show that the domain wall is practically localized within one unit cell (0.3 nm) and the magnetization displays weak oscillations around the transition place (Fig.~\ref{fig:ribbon_domain_wall}(c)). Surprisingly, we found that the collinear domain wall creation energy is only $E_{dw}=+6.5$ meV, which is more than one magnitude lower compared to case of zigzag graphene nanoribbons, $E_{dw}=+114$ meV.\cite{c_42} The strong localization and the low energy of the domain wall together indicate weak magnetic coupling along the S edge. In order to estimate the magnetic coupling, we calculate the quadratic energy-wave vector dispersion relation, $E(q)=Dq^2$ with the spin wave exchange stiffness constant, $D$. From the different $q$ vector calculations, the spin stiffness constant is found to be $D=161$ meV\AA$^2$, which is around a half compared to zigzag graphene nanoribbons, $D=320$ meV\AA$^2$.\cite{c_42} 
\par An explanation of the weak coupling in the system compared to zigzag graphene nanoribbons is related to the different geometries and electronic properties of the two materials. About the geometry, the zigzag S edge atoms distance is 3.18 \AA, larger than in the case of graphene nanoribbons' C atoms distance (2.46 \AA), which is able to reduce the interaction between the edge atoms in MoS$_2$. The nearest neighbor atoms, which can also mediate magnetic coupling between the edge atoms, are C atoms in graphene, while Mo atoms for MoS$_2$. The magnetic coupling through the middle layer Mo atom can differ from the coupling through the in-plane C atom in graphene. Besides the differences in the edge geometries, there are also discrepancies between the electronic properties. The edge states in zigzag graphene nanoribbons show almost flat bands in contrast to MoS$_2$, where the S atom bands have a small, but finite energy dispersion (Fig.~\ref{fig:ribbon_bands}). The higher density of states due to the flat bands can significantly strengthen the electron-electron interaction effects and thus the magnetic coupling in the case of graphene nanoribbons. We also verified the weak coupling by performing DFT calculations in a double unit cell geometry of zigzag MoS$_2$ nanoribbon. The states with ferromagnetic ($\uparrow\uparrow$) and antiferromagnetic ($\uparrow\downarrow$) spin ordering at the edges show only 14 meV difference in energy. It is worth noting that the small energy values between the ferromagnetic and antiferromagnetic states were reported in zigzag WS$_2$ nanoribbons,\cite{c_51} which also emphasizes the similar magnetic mechanisms in layered structures of MoS$_2$ and WS$_2$.
\subsection{Zigzag nanoribbon with disorder}
\begin{figure*}[!ht]
\includegraphics[scale=0.337]{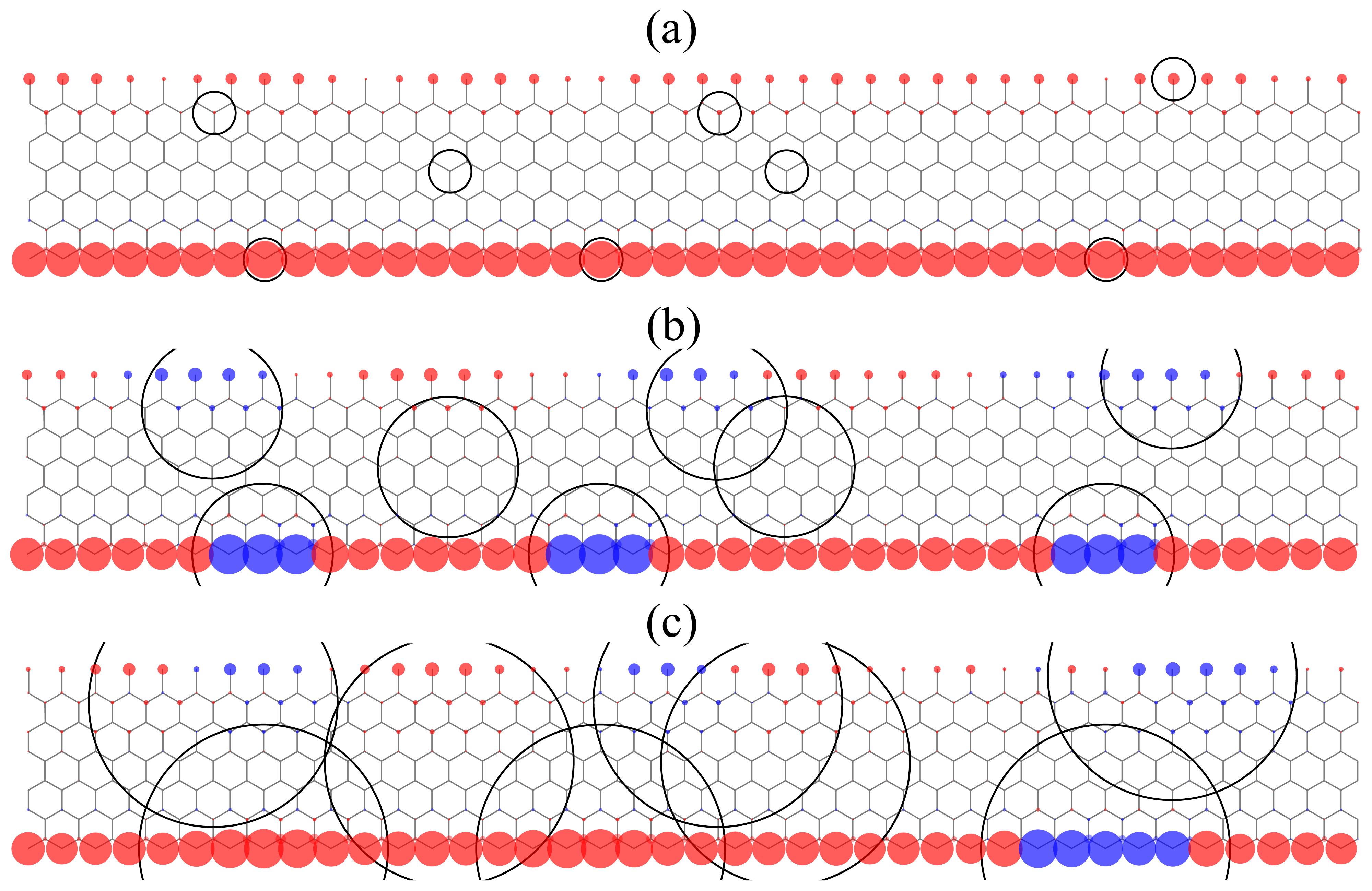}
\caption{Spin density plot in the presence of disorder. (a)-(c) Top view of the magnetic ground states of the zigzag nanoribbons by using Gaussian potentials with $V_0=+100$ meV and different width parameters $c=1,3,5$ in \AA ngstr\"om. Black circles denote the positions of the randomly distributed potentials. The radius of the circles corresponds to the size of the potentials. Blue and red circles correspond to spin-up and spin-down electrons, respectively.}
\label{fig:ribbon_disorder}
\end{figure*}
\par Defects and disorder can significantly modify the intrinsic properties of the materials. In the case of zigzag MoS$_2$ nanoribbons, transport calculations have revealed strongly suppressed conductance in the presence of edge disorder.\cite{c_40, c_41, c_52} Motivated by the observed weak magnetic coupling at the S edge, we investigate the robustness of the magnetization against short- and long-range disorder.
\par In order to model disorder in the system we apply Gaussian potentials to the on-site energy parameters on each site:
\begin{equation}
\label{eq:potential}
 V(\textbf{r})=\sum_{i=1}^{N} V_0 e^{-\lvert \textbf{r}-\textbf{r}_i \rvert^2/2c^2},
\end{equation}
where $\mathbf{r}_i$ are the positions of the disorder, $V_0$ and $c$ are Gaussian parameters corresponding to the strength and the range of the disorder, respectively. In the case of specific defects, such as vacancies or adatoms, the on-site and the hopping parameters of the TB model should be modified to describe the defect properties. However, by using the combination of our edge parametrization and Gaussian potentials, we are able to examine both short- and long-range disorder in the system without further modifications of the TB parameters. This disorder potential can be also regarded as an inhomogeneous charge distribution of the substrate.\cite{c_53}
\par We use randomly distributed potentials in the system ($N=8$), which contains disorder along the edges and also the inner part of the nanoribbon (marked by the center of the black circles in Fig.~\ref{fig:ribbon_disorder}). Gaussian potentials with $V_0=+100$ meV are considered according to often observed n-doped behavior of MoS$_2$ samples on substrates.\cite{c_54, c_55} Fig.~\ref{fig:ribbon_disorder} shows the calculated magnetic ground state of the system for $c=1,3,5$ values in \AA ngstr\"om, which corresponds to disorder localized from one atom to extended defects above nanometer size.
\par
\begin{figure}[!ht]
\includegraphics[width=\columnwidth]{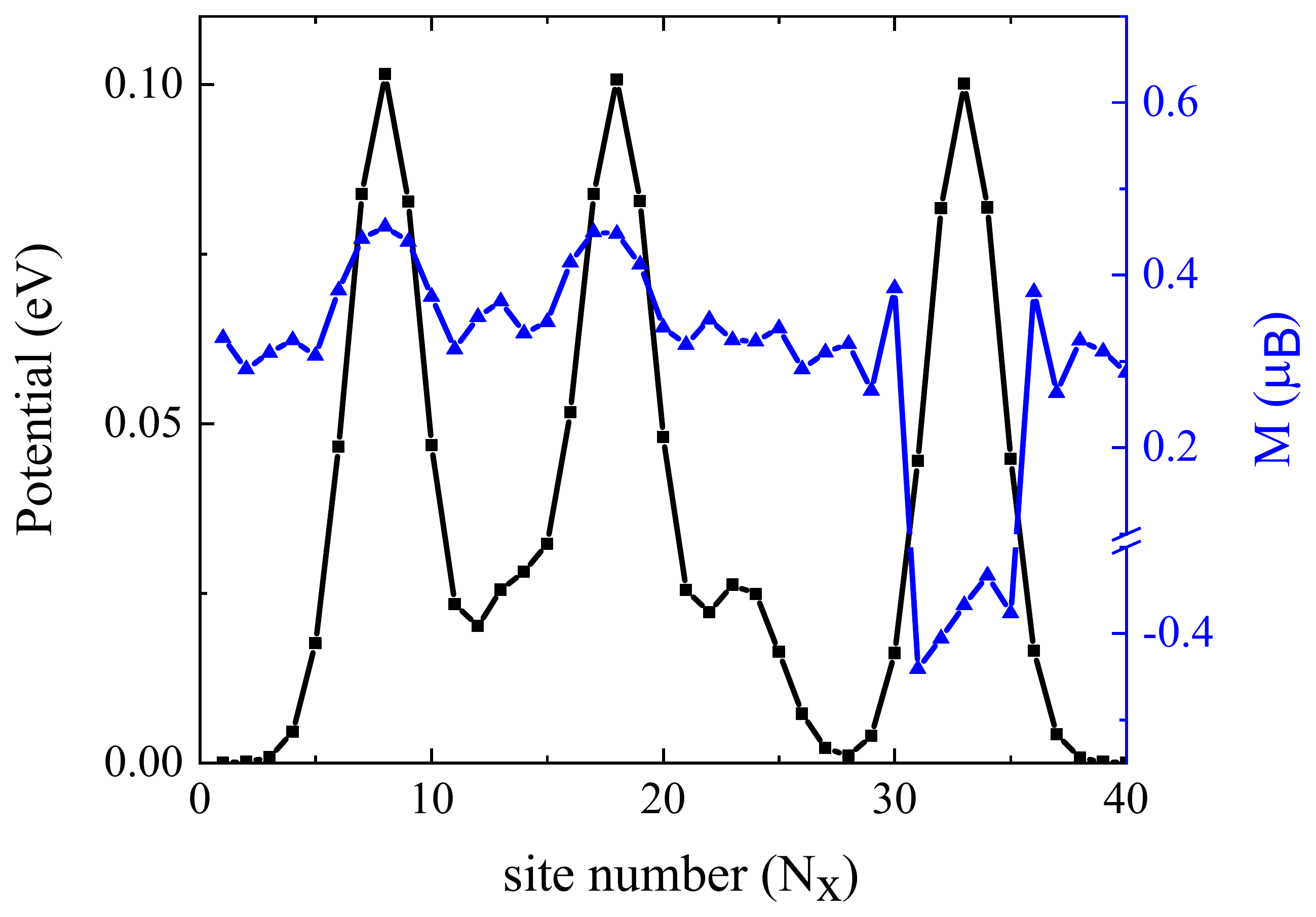}
\caption{Potential profile and magnetic moments along the S edge. Black and blue curves correspond to the potential ($c=5$ \AA) and magnetic values on the S atoms at the S edge. The magnetic moments are following the potential profile. Ferromagnetic orientation of the spins occurs, where the Gaussian potentials have overlapping region, while domain walls appear in the non-overlapping region.}
\label{fig:ribbon_potential}
\end{figure}

 In Fig.~\ref{fig:ribbon_disorder}(a), we can recognize the ferromagnetic ground state at the S edge for the case of the strongly localized perturbation potentials ($c=1$ \AA). We found that the disorder localized in the middle of the nanoribbon does not affect the magnetic properties, while the disorder on the S atoms at the S edge causes slightly increased magnetic moments from the $M=0.35$ $\mu_B$ defect-free value to $M=0.41$ $\mu_B$. The growth of the magnetic moments of the S atoms, where the potentials are centered, can be understood from Fig.~\ref{fig:ribbon_spin_bands}b. The positive potential causes positive shift of the bands in energy, therefore the partially filled spin-up band of the S atoms at the edge becomes less occupied. The spin-down band is far from the Fermi-level, therefore it remains totally occupied in the presence of the potential resulting higher magnetic moments for the S atoms. Overall, we can say that shifting the bands of the S atoms due to positive or negative potential leads to increased or decreased magnetic moments at the edges compared to the defect-free system. In contrast to the previous $c=1$ \AA~result, the potential with $c=3$ \AA~parameter extending more than 3 atom distances, cause significant changes on the magnetic ground state (Fig.~\ref{fig:ribbon_disorder}(b)). Most importantly, in regions, where the potentials are applied, the orientation of the magnetic moments has been changed  ($M=-0.47$ $\mu_B$). The lower energy of the observed state compared to the ferromagnetic state implies that potentials act in a more complex way than in the previous $c=1$ \AA~case. The potentials modify both the Mo and S atoms bands in the potential region resulting in the formation of domain walls along the edges. Similar changes of the magnetic moments can be seen at the Mo edge, where the S dimers have significantly smaller magnetic moments ($M=\pm0.05$ $\mu_B$). Further increasing of the radius of the individual Gaussian potentials ($c=5$ \AA) leads to overlapping regions in the potential at the edges (Fig.~\ref{fig:ribbon_disorder}c). The magnetic calculations reveal that in the case of the overlapping region the ferromagnetic ground state is restored at the S edge. In contrast, domain walls appear in non-overlapping region in the S edge. The relation between the potential and magnetic texture is visible in Fig.~\ref{fig:ribbon_potential}. This result implies that besides the width, the profile of the potential plays also an important role on the magnetic ground state of the system. 
\par In conclusion, our calculations including disorder highlight significant changes on magnetic ground state in zigzag MoS$_2$ nanoribbons. For short range disorder the values of the magnetic moments have been changed, but the ferromagnetic arrangement is still conserved. By increasing the disorder range not only the values, but also the direction of the moments is modified yielding spin domain walls, which are also sensitive to the profile of the potential. The energy differences of the ground states and excited states in the different disorder are in the order of tens meV in the system. This behavior can be qualitatively understood if we assume that the edge magnetic moments can be described by a one-dimensional Ising model, as it was shown for graphene with zigzag edges.\cite{c_42} In this case, the effects of the disorder resemble to what the random fields cause in the Ising model, where also formation of domain walls in the system was predicted.\cite{c_56, c_57, c_58} In those systems, the creation of the domain wall is determined by an interplay of the domain wall energy and the properties of the applied field. It seems that in our case the properties of the disorder potential play a similar role. The possibility to modify the spin-texture by potential disorder can be useful for spintronic applications. By applying periodic or non-periodic potentials at the edge, the magnetic moments can be tuned realizing various magnetic ground states. Furthermore, even dynamical control of the edge magnetic moments can be achieved with local probe techniques (eg. conducting AFM tip), where the tip induced potential flips the edge moments at the location of the tip. By moving the tip along the edge, one could move the induced reversed magnetic domain.
\section{Conclusions} 
\par Magnetic properties of 2D materials continue to pose a great interest both from fundamental and application point of view. In this paper, we reported magnetic calculations for MoS$_2$ nanoribbons with zigzag edges based on the Hubbard model. We demonstrated that proper TB parametrization of the edge atoms is crucial in order to describe the magnetic properties of the nanoribbons. Using our fine-tuned TB parameters and Hubbard interaction strength, we have investigated a several nanometer long ribbon and calculated the spin domain-wall energy. The observed low domain-wall energy indicates weak magnetic coupling between the S atoms at the edge in contrast to zigzag graphene nanoribbons. By using randomly distributed Gaussian potentials we have also revealed the effect of the disorder on the magnetic properties. We have shown that the magnetic ground state strongly depends on the potential parameters, where even disorder with few atomic distances can change the orientation of the edge spins. While these findings reveal the importance of reducing the disorder in MoS$_2$ nanoribbons (for example by encapsulating the ribbons with hexagonal boron nitride), this feature can be also exploited to manipulate the spin-texture by an applied potential field. Our approach presented here opens the way for investigating electron-electron effects in large scale MoS$_2$ and other TMD materials, which is essential for spintronic applications.
\acknowledgements{P.~V.~and I.~H.~contributed equally to this work. The work has been supported by the NanoFab2D ERC Starting Grant, the Graphene Flagship, H2020 Graphene Core2 project no.~785219, the H2020 MCA-RISE "Infusion" project no.~734834 and the Korea Hungary Joint Laboratory for Nanosciences. P.~V. acknowledges the Hungarian National Research, Development and Innovation Office (Hungary) Grant No.~KH130413. I.~H.~was supported by the Alexander von Humboldt Foundation and in part by Hungarian National Research, Development and Innovation Office through Grant No.~K120569 and the Hungarian Quantum Technology National Excellence Program (Project No.~2017-1.2.1-NKP-2017-00001). The research used resources of the “Plateforme Technologique de Calcul Intensif (PTCI)” located at the University of Namur, which is supported by the FNRS-FRFC under Conventions No.~2.4520.11.}

\bibliography{paper_MoS2}

\end{document}